\documentclass[times, twocolumn]{IEEEconf} 

\usepackage[colorlinks]{hyperref}
\usepackage{breakurl}
\usepackage{listings}
\usepackage{graphicx}
\usepackage{amstext}
\usepackage{bbm}
\usepackage{amsmath}
\usepackage{amsfonts}
\usepackage{amssymb}
\usepackage{xspace}
\usepackage{multirow}

\renewcommand{\url}{\path}

\usepackage{color}

\newcommand{\eg}{e.g.\xspace}

\newcommand{\etc}{etc.\xspace}

\newcommand{\apriori}{\emph{a priori}\xspace}

\expandafter\def\expandafter\UrlBreaks\expandafter{\UrlBreaks \do\*\do\-\do\~\do\'\do\"\do\-}

\author{Konstantin Berlin\\
	\begin{affiliation}
		Invincea Labs, LLC 
	\end{affiliation}\\
	\email{kberlin@invincea.com}
\and
Joshua Saxe\\
	\begin{affiliation}
		Invincea Labs, LLC 
	\end{affiliation}\\
	\email{josh.saxe@invincea.com}
}
\begin{document}

\title{Improving Zero-Day Malware Testing Methodology Using Statistically Significant Time-Lagged Test Samples}



\maketitle

\begin{abstract}

Enterprise networks are in constant danger of being breached by cyber-attackers, but making the decision about what security tools to deploy to mitigate this risk requires carefully designed evaluation of security products. One of the most important metrics for a protection product is how well it is able to stop malware, specifically on ``zero''-day malware that has not been seen by the security community before. However, evaluating zero-day performance is difficult, because of larger number of previously unseen samples that are needed to properly measure the true and false positive rate, and the challenges involved in accurately labeling these samples. This paper addresses these issues from a statistical and practical perspective.  Our contributions include first showing that the number of benign files needed for proper evaluation is on the order of a millions, and the number of malware samples needed is on the order of tens of thousands. We then propose and justify a time-delay method for easily collecting large number of previously unseen, but labeled, samples. This enables cheap and accurate evaluation of zero-day true and false positive rates. Finally, we propose a more fine-grain labeling of the malware/benignware in order to better model the heterogeneous distribution of files on various networks.

\end{abstract}

\section{Introduction}

Anti-malware product vendors often submit their security products to independent malware testing organizations to provide an unbiased evaluation of their products' performance \cite{avtest2016,nss2016}. Properly done, these tests provides a critical evaluation of the product that can help guide purchasing decisions for individuals and organizations. One of the critical aspects of the tests is figuring out the performance of malware detection products on ``zero''-day malware. Here we use the term zero-day to mean malware previously unseen by the computer security community, rather than the more specific definition describing malware that utilized a previously unknown exploit \cite{zeroday}.

The typical malware protection product consists of several major detection components; among them are the malware static detection engine (signatures and/or machine-learning), (heuristic) behavioral detection engine, and a reputation (or intelligence) engine. \cite{symantec2016}. Static/dynamic detection engines, which are typically deployed on endpoints and are used to make on-the-fly malware blocking decisions, are conceptually different from reputation engines, which are deployed on remote servers and inform administrators of infections after the fact\cite{virustotal2016}\cite{polonium2011}. Today's reputation engines usually consist of software whitelists and blacklists, and are thus not relevant to the problem of detecting zero-day malware; we thus leave aside discussion of the evaluation of anti-malware reputation engines in this paper.

This paper's contributions pertain to evaluating zero-day performance, which is important in understanding how well a product will perform against malware used in targeted attacks. We note that this is not the only performance metric for a detector, and it is also reasonable to measure detector's performance on common in-the-wild malware, which more closely resembles home user's experience when browsing the internet.\footnote{Typically, when anti-malware products performance is reported, the focus is on the detection rate, also referred to as the true positive rate (TPR); simply defined as the fraction of all malware that was detected. However, measuring TPR cannot possibly provide an accurate picture of the detector, since it gives no information about the false positive rate (FPR). 

In practice, during demonstrations, one could simply set the detector in a more aggressive setting than can be practically deployed, in order to get an inflated detection score. During deployment, the number of benign files on a system/network far outnumber the malware files, thus detector's performance on the benign files contributes significantly more to the overall detection performance of the system than TPR. Since it is typically unclear what the ratio of malware to benignware is on the system \apriori, we focus our evaluation on TPR and FPR metrics (as opposed to precision and recall), which are independent of this ratio.}

There are three main challenges in evaluating a malware detection system: i) the large class imbalance in the environment, meaning that benign binaries are vastly more common than malicious binaries;  ii) the difficulty in getting legitimate benign and malware samples that truly have never been seen by the anti-virus industry prior to testing; and iii) the difference in the malware/benignware mix between the testing environment and actual deployment of the system,

Below we address the above three issues by first deriving the number of malware and benignware samples that are needed to perform statistically rigorous test. We derive the proper sample size that should be used for evaluation, and demonstrate the issue of bias when the sample size is too small. Then, we propose a time-delay test to separate the performance of the detector from the reputation system. Finally, we briefly describe the idea of hierarchical testing in order to evaluate performance on various sets of malware/benignware distributions that might better reflect the heterogeneous environments where the AV detection system might be deployed.

\section{Estimating Proper Sample Size}

To provide valid detection rates for a given AV detector, enough samples must be collected to render the results statistically significant \cite{nist2016}. The accuracy of the observed FPR and TPR estimates are related to the sample size $N$ used to estimate them, as well as the actual true rate for either FPR and TPR. The confidence bounds for the estimated rate can be accurately determined by observing that predicting if a file is benignware or malware is akin to flipping a weighted coin, where the weight $p$ is the true rate. Given $N$ flips of the coin, $\hat{p}$, the unbiased estimator for $p$, is computed by taking the number of positive guesses and dividing by $N$ \cite{brown2001interval}. The distribution for the expected number of correct malware labels (need to compute $\hat{p}$) given $N$ samples is determined by the binomial distribution.

Ford and Cavalho \cite{ford2014significant} provided an interesting discussion on the importance of sample size $N$ in malware testing. Here we expend on this discussion by asking a practical question, how many samples do we actually need when designing a malware detection test? We answer this question by doing the inverse analysis of \cite{ford2014significant}, where rather than asking, given the sample size, what are the confidence internals, we determine, given the FPR or TPR, what is the proper samples size.

We start with the known solution to the forward problem, the probability that the estimate is correct given the sample size, and then use global optimization to solve for the sample size, given the desired confidence probability. Specifically, we want to compute the sample size $N$ that would give us at least $c$ confidence (we will use $c=0.95$ for 95\% confidence) that our estimated FPR or TPR $\hat{p}$ is within $\sigma$ of the true value $p$. To compute this value, we first find the probability of such an event occurring, and then solve for $N$ given the desired $c$.

The probability of the estimate being approximately correct, $c$, is
\begin{equation}
	\label{eq:bound}
	Pr(p-\sigma < \hat{p} < p+\sigma) = F_p(p+\sigma)-F_p(p-\sigma),
\end{equation}
where $F_p$ is the cumulative distribution function (CDF) of possible outcomes for $\hat{p}$. In our case $\hat{p}$ is the ratio of the observed positive samples divided by the number of observed samples, and so $F_p$ can be expressed using the binomial distribution's CDF $G_p$, $F_p(x) = G_p(N x)$. 

\begin{figure*}[ht]
	\center
	\includegraphics[width=.8\textwidth]{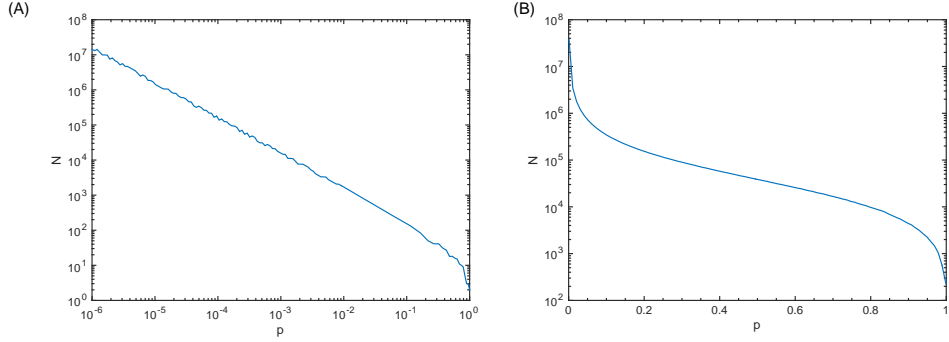}
   \caption{ The estimated sample size needed to accurately (95\% confidence) compute the desired rate within a certain percent $\alpha$, $p-p\alpha<\hat{p}<p+p\alpha$. (A) $N$ for a range of values useful when estimating false positive rates, with $\alpha=0.5$. (B) $N$ for a range of values using when estimating the true positive rates, with $\alpha=0.01$. }
   	\label{fig:size}
\end{figure*}

Given a desired $p$ value, we solve eq. \eqref{eq:bound} for $N$, using a global optimizer or simply trying every possible value of $N$ within a reasonable range. We show $N$ needed to accurate estimate for a range of reasonable values of $p$ in Fig. \ref{fig:size}.

Fig. \ref{fig:size} shows that that for accurate FPR estimates (where typical $p \in [10^{-5}, 10^{-3}]$)  we need about a million benign samples, while to accurately estimate TPR (where typically $p \in [0.5,0.95]$) we need about 10 thousand samples.

\begin{figure*}[ht]
	\center
	\includegraphics[width=1.00\textwidth]{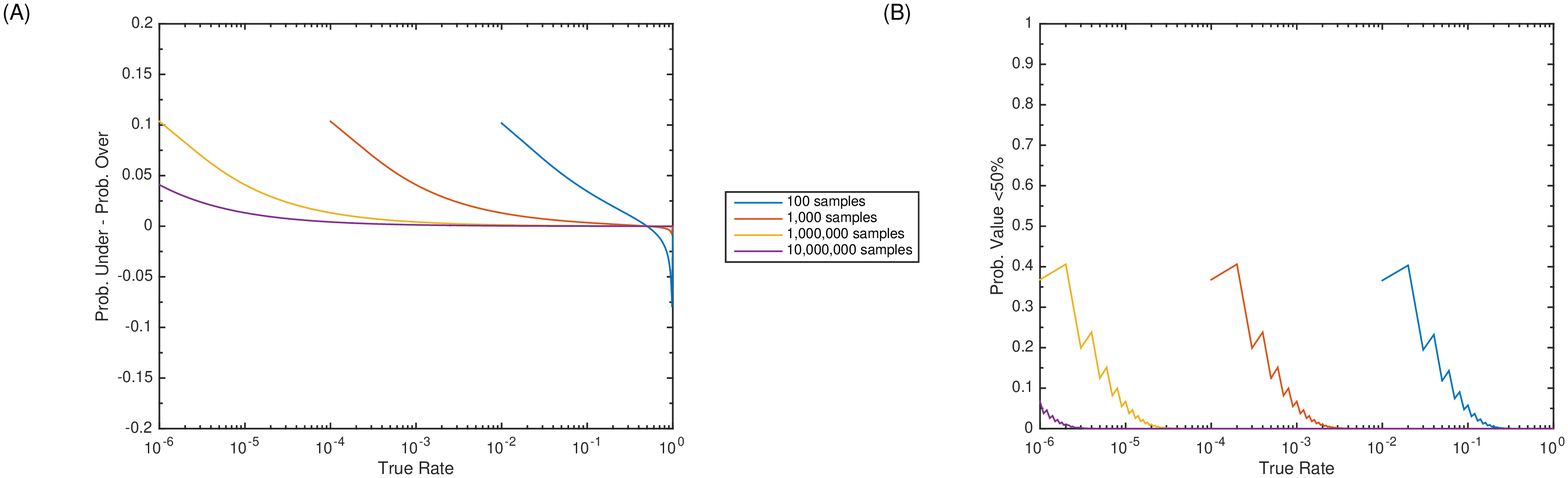}
   \caption{ The probability of underestimating the TPR of FPR rates given several common samples sizes. (A) The probability that the estimate $\hat{p}$ will be biased towards underestimation, given the true rate $p$. (B) The probability that the rate estimate $\hat{p}$ will be more than 50\% smaller than true rate $p$. }
   	\label{fig:bias}
\end{figure*}

In today's security industry, small sampling is common in impromptu point of sale evaluations, and is even found in testing done by major testing services, because getting a large number of realistic malicious and benign samples is expensive and time consuming. Interestingly, in small, informal sample testing regimes the estimated rates can be biased compared to the true rates. We demonstrate this by plotting $F_p(p-\sigma-\epsilon)-(1-F_p(p+\sigma)$, probability of underestimated the true rate minus probability of overestimating; and $F_p(p-0.05p)$, the probability of underestimating the rate by over $50\%$, in Fig. \ref{fig:bias}. 

Fig. \ref{fig:bias}A,B demonstrate two interesting properties of binomial distribution. Fig. \ref{fig:bias}A shows that while $\hat{p}$ is an unbiased estimator of $p$, the probability of underestimating the score is higher than probability of overestimating the score when $p$ is small (such as for FPR), and the reverse when $p$ is large (such as for TPR). This observation is discussed in further detail in \cite{brown2002confidence}. Fig. \ref{fig:bias}B shows that the probability of significantly underestimating the rate (by 50\% or more), also gets much larger as sample size gets small. The non-continuous lines in Fig. \ref{fig:bias}B are due to the discrete nature of the binomial distribution, and has been discussed at length in \cite{brown2001interval}. 

Our observations above have two immediate implications: i) if the testing sample size is unreasonably small for TPR evaluation (where typically $p \in [0.1,1.0]$), $<1000$ samples, the evaluation would tend to overestimate the detection rate, making the detector look better than the true performance; and ii) if the testing samples size is too small for FPR estimation (where $p \in [10^{-5}, 10^{-3}]$), $<10^{5}$ samples, the evaluation would produce an overly optimistic FPR rate. 

We confirm the above theoretical analysis by performing a validation test on our previously developed malware detector \cite{saxe2015deep}. We use 600,000 samples, approximately evenly split between benign and malware, as well as larger 1.3 million sample data-set to compute the ROC curve of our detector. The ROC curve is simply a set of individual (TPR, FPR) pairs, and so we would expect the FPR bias to appear at the lower range of the curve. 

\begin{figure}[ht]
	\center
	\includegraphics[width=.45\textwidth]{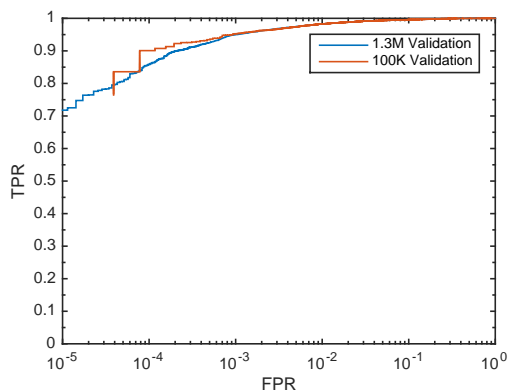}
   \caption{ The ROC curve of true positive rate (TPR) vs. false positive rate (FPR) of a machine-learning based detector. The blue line shows validation using 1.3 million file. The red line shows validation using 100 thousand files. The overly optimistic bias for the red line is due to optimistic FPR estimates, as theoretically predicted. }
   	\label{fig:perf}
\end{figure}

Indeed, as the results in Fig. \ref{fig:perf} show, the 600,000 sample validation line (red line) is skewing the FPR rate higher in the low FPR region, resulting in falsely optimistic ROC curve.
 
\section{Getting New Samples Using Time Delays}

The results shown above show that a million files are needed to properly evaluate the FPR of a detector. How do you get that many previously unseen labeled files? The traditional approach from the testing service has been to use a data feed to get a supply of malware and benignware. However, there are several problems with this approach. The primary issue is that a majority of these samples are not new, and can already be found in large malware feed networks like VirusTotal \cite{virustotal2016}. Furthermore, large vendors have vast deployments that allow them to potentially observed most files even before they are uploaded to VirusTotal. For example, evaluation in \cite{kwon2015dropper} suggested that large vendor AV-systems can observe new malware over 100 days before they show in VirusTotal malware feed. This suggests that large fraction of files that are traditionally used for evaluation can be blacklisted, whitelisted, or used for authoring signatures and training machine learning models \apriori, skewing the results of tests focused on the zero-day detection problem.

One naive approach to mitigating this problem would be to generate the malware samples in-house. But this is not feasible for millions of samples that are required for proper evaluation, and would produce an unrealistic test set compared to what is being observed in the wild. Another mitigation would be disconnecting the test environment from the Internet, in order to prevent anti-malware products from communicating with their reputation servers \cite{nss2016}, but that doesn’t account for extensive white-listing that can be done on the endpoint beforehand.

The second major issue involved in testing anti-malware products against zero-day malware are the inaccurate labels that we will inevitably have for new files, since most vendors used to generate the labels might not have caught up with the new samples.  We propose to mitigate the issue of blacklist/whitelists, as well as the incorrect labels skewing the results by proposing a time delayed experiment, where the detector is disconnected from the updating service for several months (\eg 100 days). Once the time delay expires, newly unseen files are aggregated from a file feed and fed through the detector. The detection statistic is evaluated on this set of files.

\section{Hierarchical Evaluation Statistics}

Assuming the evaluation is statistically valid, the other major problem when evaluating AV products is the need for accurate modeling of the malware and benign distributions. Any detector is designed to balance the chance of false positives with the chance of a detection. However, only some subset of malware/benignware will actually be observed on any given network. The ability to detect malware that the users will never see is irrelevant to that user's experience. However, it is not really possible to know what a realistic distribution of malware/benignware is for a user, and very hard to even generalize for a prototypical enterprise. On the other hand, the major industry players know fairly well what to expect, due to their large deployment on endpoints. Therefore, we propose a hierarchical evaluation that would allow flexible recomputation of the detector for specific deployment environments.

The hierarchical process consists of two phases. In the first phase, the benign and malware files will be broken down into more fine-grained hierarchical categories, like common benign files (\eg files prom popular Microsoft, Adobe products, \etc), shareware/adware files from download websites, business productivity software. Separating the files into multiple subcategories allows us to recompute the expected performance of a detector by recombining/weighing the the data to match the more specific file distributions you would expect to see on various networks.

For example, the variety of software installed on an enterprise network is more limited than you would see on the same number of individual computers. At the same time, the amount of malware/adware that you would likely to see on an individual computer is higher. Therefore, when evaluating the detectors performance, as it relates to enterprise vs. individual computers, the enterprise evaluation should weigh benignware more than it would for individual evaluation. However, the scope of the false positive testing should be limited to more traditional enterprise software rather than less software found on free download websites.

Given these categories, anti-malware product vendors can specify their own mix of categories that they deem proper for evaluation of their engine, and direct comparison can be made between the vendors by taking a weighted sum of the FPRs and TPRs to match a desired distribution.

\section{Conclusion}

In this paper we have made three key contributions.  First, we have demonstrated the importance of proper sample size in estimating the true and false positive rates of malware detectors. The sample size required, specially for estimating the false positive rates is significantly higher than commonly used in testing services, and requires on the order of a million files. While the requirement to estimate the true positive is lower, it still needs thousands of samples to accurately estimate.

Second, to address the problem of creating a large test set for evaluating zero-day malware detection, we have proposed a time-delay test, where the detectors under test are simply isolated for several months, while new samples are being gathered. This proposed time-delay test potentially provides a more accurate evaluation of the zero-day performance than current tests. 

Third, to further refine the evaluation, we have proposed individually testing a subcategories of samples, and then aggregating them to provide more realistic performance under various user environments.

We hope that the above analysis provides a basis for improvement in anti-malware testing methodologies, allowing for more accurate evaluation of performance. 

\sloppy

\bibliographystyle{abbrv}
\bibliography{main}

\end{document}